\documentclass[twocolumn,preprintnumbers,amsmath,amssymb]{revtex4}
\usepackage{graphicx}  
\begin{document}
%%%%%%%%%%%%%%%%%%%%%%%%%%%%%%%%%%%%%%%%%%%%%%%%%%%(one column)
%\documentclass[manuscript]{revtex4}
%\newcommand{\MF}{{\large{\manual META}\-{\manual FONT}}}
%\newcommand{\manual}{rm}       
%\newcommand\bs{\char '134 }    
% 
%\usepackage{graphicx}  
%\begin{document}
%%%%%%%%%%%%%%%%%%%%%%%%%%%%%%%%%%%%%%%%%%%%%%%%%%%%
\title{Graphene terahertz uncooled bolometers
}
\author{V.~Ryzhii$^{1,3}$, T.~ Otsuji$^{1}$, M.~Ryzhii$^{2}$,  N.~Ryabova$^3$, 
S.~ O.~Yurchenko$^3$, V.~Mitin$^4$, 
and M.~S.~ Shur$^5$
}
\affiliation{
$^1$ Research Institute for Electrical Communication,
Tohoku University, Komada,  Sendai 980-8577, Japan\\
$^2$Computational Nanoelectronics Laboratory, University of Aizu, 
Aizu-Wakamatsu 965-8580, Japan\\
$^3$ Center for Photonics and Infrared Engineering, Bauman Moscow State Technical University,  Moscow  105005, Russia,\\
$^4$ Department of Electrical Engineering, University at Buffalo, Buffalo,
NY 1460-1920, USA,\\
$^5$ Department of Electrical, Electronics, and System Engineering, Rensselaer Polytechnic Institute, Troy, NY 12180, USA.  
}

\begin{abstract}
We propose the concept of  a  terahertz (THz) uncooled  bolometer based on
n-type and p-type  graphene layers (GLs), constituting the absorbing regions,  connected by an array of 
undoped graphene nanoribbons (GNRs).
The GLs absorb the THz radiation   
with the GNR array playing the role of the barrier region (resulting in nGL-GNR-pGL bolometer).
The absorption of the incident THz radiation in the GL n- and p- regions leads to variations of the 
effective temperature of electrons and holes and of their Fermi energy resulting in the variation 
of the current through the GNRs. Using the proposed device model, we calculate the dark current and the 
bolometer responsivity as  functions of the GNR energy gap, applied voltage, and the  THz frequency.
We demonstrate that the proposed bolometer can surpass the hot-electron bolometers  using traditional 
semiconductor heterostructures.
 \end{abstract}

\maketitle

\section{Introduction}

\vspace*{-2mm}

 Owing to the gapless energy spectrum~\cite{1}, graphene layers (GLs)  absorb electromagnetic radiation  
in a wide spectral range (from the ultraviolet to terahertz)
due to the interband transitions~\cite{2,3,4}. Therefore,
GLs  can be used in   photodetectors, light sources,  modulators, and mixers using the interband transitions~\cite{5,6,7,8,9,10,11,12,13,14,15,16,17}.
The performance of these devices can be enhanced by utilizing multiple-GL structures~\cite{18}.
For the  infrared and visible spectral ranges,  the interband absorption prevails over
the intraband (Drude) absorption. However,  in  the terahertz (THz) range, especially at  low THz frequencies, the Drude absorption can dominate.
The intraband absorption in GLs  can also be used in different devices for THz modulation
and  detection. The THz detectors, including  uncooled detectors,  exploiting the effect of electron or hole heating (hot-electron or hot-hole bolometers) in
two-dimensional electron (hole) heterostructures made of
 A$_3$B$_5$, CdHgTe, and other  compound systems  were realized
previously~\cite{19,20,21,22,23}. 
In this paper, we propose and analyze THz uncooled bolometric detectors  based on  GL 
structures. We demonstrate that such bolometers  can exhibit
fairly high responsivity, effectively operating at room temperatures and surpassing THz bolometers 
based on the traditional semiconductor heterostructures. The main advantages of GL-based room temperature bolometers are associated with the following three factors: (i) high electron and hole THz  conductivities at room temperature~\cite{1} and, hence, elevated Drude absorption; (ii) the dominant mechanism
establishing   the interband and intraband equilibrium is the interaction with optical phonons~\cite{24,25};
(iii) long time of the electron and hole energy relaxation via optical phonons due to their large  energy $\hbar\omega_0 \simeq 200$~meV~\cite{1} (this time  is proportional to a factor
$\exp(\hbar\omega_0/T_0)$ and  is very large for GLs even at room  temperature $T_0 = 300$~K).

\section{Model and main equations} 
%%%%%%%%%%%%%%%%%%%
Figures 1(a) and 1(b) show the  proposed  nGL-GNR-pGL bolometers.
The bolometers consist of two gapless  n-type and p-type GL absorbing regions connected by an undoped array of GNRs with sufficiently large 
energy gap $\Delta$ (serving as the barrier region).  The GLs can be doped  chemically
[as in Fig.~1(a)] or "electrically" (using the conducting gates with 
the bias voltages, $\pm V_g$,  of different polarity, as shown in Fig.~1(b)). The gates which control the electron and hole densities can
be made using   GLs~\cite{17,26,27}.
It is assumed that the GNR width, $d$,  is sufficiently small,  so that
 the energy gap $\Delta \propto v_W/d$, (where $v_W \simeq 10^8$~cm/s is the characteristic velocity of electrons and holes in GLs)
 is large enough to provide essential confinement of electrons in the n-GL 
 and holes in the p-GL due to the formation of the barrier. The
 room temperature operation of field-effect transistors with sub 10~nm  GNRs exhibiting fairly large energy gap
 was reported in Ref.~\cite{28}. The energy barrier in such GNRs ensures a relatively strong dependence of the 
current on the effective temperature of electrons and holes enhancing  the bolometer responsivity.

\begin{figure}[t]
\vspace*{-0.4cm}
\begin{center}
\includegraphics[width=7.0cm]{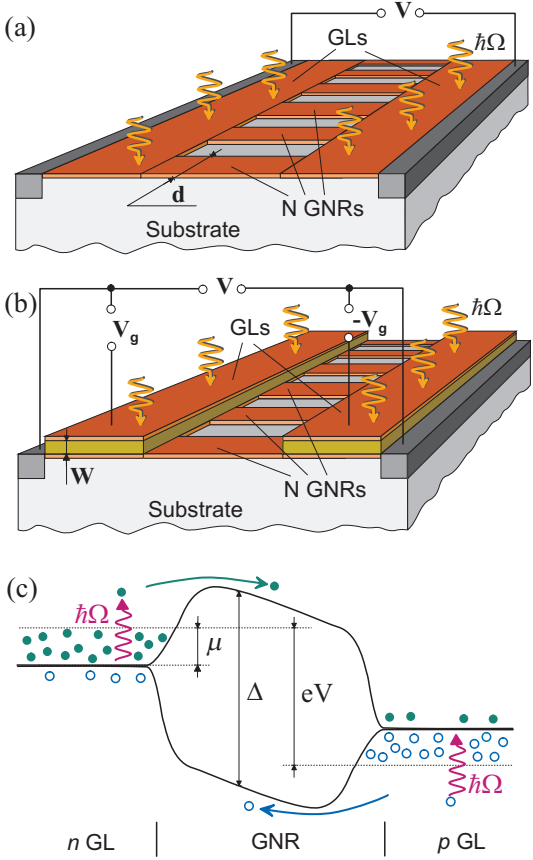}
\caption{Schematic views of  bolometers under consideration with (a) chemically doped GLs,
(b) electrically doped GLs, and (c) the bolometer  energy diagram under  bias voltage $V$ 
  (wavy arrows correspond to 
intraband transitions due to absorption of photons in GLs, smooth arrows indicate propagation of electrons and holes above
the pertinent barriers in GNRs). 
}
\end{center}
\end{figure}

Figure~1(c) shows
the resulting device band structure at sufficiently large bias voltage 
$V > - V_{bi} = 2\mu_0/e$, where $V_{bi}$ is the built-in voltage, $\mu_0$ is the Fermi energy of electrons  and holes  in GLs in equilibrium, and $e$ is the electron charge.

In the following, we assume that the interband absorption is relatively weak in comparison with the intraband absorption.
This occurs when
 the energy of photons, $\hbar\Omega$, of the incident THz radiation
is relatively small (corresponding to the frequency about few THz and lower). If $\hbar\Omega < 2\mu_0 < \Delta$, the interband transitions are forbidden  due to the Pauli blocking.
We assume that due to relatively high electron and hole densities,
 the intercarrier scattering time is sufficiently  short to provide fast Maxwellization (or Fermization)
of the photoexcited
electrons and holes. Therefore, the electron and hole systems in GLs are characterized by
quasi-Fermi energy $\mu$  and by the effective temperature $T$.
The heating of the electron and hole gases in the pertinent sections, i.e., the deviation of the effective temperature $T$ from the lattice temperature $T_0$ leads
to the deviation of the Fermi energy $\mu$ from its equilibrium (dark) value $\mu_0$. 
The quantities $\mu$ and $T$ are related by the following equation:
 $$
 \int_0^{\infty}\frac{d\varepsilon\varepsilon}
 {1 + \displaystyle\exp\biggl(\frac{\varepsilon - \mu}{T}\biggr)}
  $$
  \begin{equation}\label{eq1}
  -\int_0^{\infty}\frac{d\varepsilon\varepsilon}
 {1 + \displaystyle\exp\biggl(\frac{\varepsilon +\mu}{T}\biggr)} = \frac{\pi}{2}\hbar^2v_W^2\Sigma_0.
\end{equation}
In the case of chemical doping, the quantity $\Sigma_0$ is equal to the donor (acceptor) density. In the detectors with electric doping, $\Sigma_0$ is given by
$ \Sigma_0 = (kV_g/4\pi\,eW)$, so that $\mu_0 = \hbar\,v_W\sqrt{kV_g/4eW}$,
where $k$ and $W$ are the dielectric constant and the thickness of the layer separating GLs and the gates
and $V_g$ is the gate voltage [see Fig.~1(b)].
In the case under consideration,  the electron and hole systems are sufficiently strongly degenerated
 ($\mu_0 \gg T_0$), hence,  the Fermi energy is given by
$\mu_0 \simeq  \hbar\,v_W\sqrt{\pi\Sigma_0}$.

Considering the one-dimensional electron and hole transport in GNRs and the Fermi distributions
of electrons and holes in GLs, in particular, near the GNR edges at $V >  2\mu/e$,
the sum of the electron and hole currents (i.e., the terminal current) between the p- and n-regions
through $N$ parallel GNRs is equal to

$$
J = \frac{8eN}
{2\pi\hbar}\int_{\Delta/2}^{\infty} d\varepsilon_{GNR}
\biggl\{
\biggl[\exp\biggl(\frac{\varepsilon_{GNR} - \mu}{T}\biggr) + 1\biggr]^{-1}
$$
\begin{equation}\label{eq2}
-
\exp\biggl[\biggl(\frac{\varepsilon_{GNR} - \mu - eV}{T}\biggr) + 1\biggr]^{-1}
\biggr\}.
\end{equation}
Here 
$\varepsilon_{GNR}$  is the kinetic energy of  electrons and holes in GNR.
In the absence of illumination, 
i.e., when $\mu = \mu_0$ and $T = T_0$, Eq.~(2) yields
the following expression  for the dark current $J_{0}$:

\begin{equation}\label{eq3}
J_{0} \simeq \frac{4eT_0N}
{\pi\hbar}\ln\biggl[\exp\biggl(\frac{\mu_0 -\Delta/2}{T_0}\biggr) + 1\biggr].
\end{equation}
%
%%%%%%%%%%%%%%%%%%%%%%%%%%%%%%%%%%%%%%%%%%%%%%%????????????????????
Setting  $\Delta/2 - \mu_0 = 25$~meV, 
 and $N = 1$, for $T_0 = 300$~K  we obtain
$J_0 \simeq 2.64~\mu$A. This value is in a good agreement with experimental results~\cite{28}.

%%%%%%%%%%%%%%%%%%%%%%%%%%%%%%%%%%%%%%%%%%%?????????????????

At relatively weak irradiation,  $|T - T_0| \ll T_0$ and $|\mu - \mu_0| \ll T_0$.
Considering this,
the variation of the current through the GNR array, $(J - J_0)$,  i.e., the photocurrent, 
can be presented in the following form:
$$
J - J_{0}\simeq \frac{4eT_0N}
{\pi\hbar}\frac{\displaystyle\exp\biggl(\frac{\mu_0 -\Delta/2}{T_0}\biggr)}{\biggl[\displaystyle\exp\biggl(\frac{\mu_0 -\Delta/2}{T_0}\biggr) + 1\biggr]}
$$
\begin{equation}\label{eq4}
\times\biggl[\biggl(\frac{\Delta/2 - \mu_ 0}{T_0} + 1\biggr)\frac{(T - T_0)}{T_0}  
+ \frac{(\mu - \mu_0)}{T_0}\biggr].
\end{equation}
The first and the second terms in the right-hand side of Eq.~(4) describe the effect of variation of the effective temperature and the quasi-Fermi energy due to heating by the THz radiation. However, as follows from Eq.~(1), when  $\mu_0 \gg T_0$,
the variation of the quasi-Fermi energy is relatively small,
hence, the last term in the right-hand side of Eq.~(4) can be omitted. 
Considering that 
the energy relaxation due  to the processes
  governed by the interaction with optical phonons, the electron and hole  effective temperature $T$  and the number of optical phonons ${\cal N}_0$
obey the following  equations:

\begin{equation}\label{eq5}
R_0^{intra}   = R^{decay}, 
\end{equation}
\begin{equation}\label{eq6}
 \hbar\omega_0\, R^{decay} = \hbar\Omega\,I_{\Omega}\beta\,g_{\Omega}^{intra}.
\end{equation}
Here, $I_{\Omega}$ is the THz photon flux, $\beta = \pi\alpha$, where $\alpha = e^2/c\hbar$, $e$ is the electron charge, $c$ is the speed of light, 
 $R_0^{intra} = R_0^{intra}(T,  {\cal N}_0)$ is the rate of 
the  intraband 
transitions  involving the emission and absorption of optical phonons,
 $R^{decay} = R^{decay}({\cal N}_0)$ is the rate of optical phonon decay, and  $g_{\Omega}^{intra}$ 
is proportional to the GL Drude ac conductivity ~\cite{29,30}:
$$
g_{\Omega}^{intra} =
\frac{4T_0\tau}{\pi\hbar(1 + \Omega^2\tau^2)}
\ln\biggl[\exp\biggl(\frac{\mu_0}{T_0}\biggr) + 1\biggr]
$$
\begin{equation}\label{eq7}
\simeq \frac{4\mu_0\tau}{\pi\hbar(1 + \Omega^2\tau^2)},
\end{equation}
where $\tau$ is the  momentum relaxation time of electrons and holes, which, generally,
is depending
on  $\mu_0$ and $T_0$.
Equations~(5) and (6) govern 
the balance of the energy of the electron-hole system and
  the number optical phonons in GLs explicitly accounting for
 all the energy
received by the electron-hole-optical phonon system from THz radiation going eventually to the thermostat.

Since $\mu_0, T_0 \ll \hbar\omega_0$, the expression
 for the term $R_0^{intra}$ can be
simplified ~\cite{31}:

\begin{equation}\label{eq8}
R_0^{intra} =  \frac{\Sigma_0}{\tau_0^{intra}}
\biggl[({\cal N}_0 + 1)\exp\biggl(-\frac{\hbar\omega_0}{T}\biggr) 
- {\cal N}_0\biggr]. 
\end{equation}
Here, 
 $\tau_0^{intra} $  is the time 
of the   intraband  
phonon-assisted processes: the quantity
$\tau_0^{intra} /{\cal N}_0\simeq \tau_0^{intra}\exp(\hbar\omega_0/T_0)$ 
plays the role of the effective  energy relaxation time of electrons and holes. 
%Their ratio  can be approximated as~\cite{31,33}
%$(\tau_0^{intra}/\tau_0^{inter}) \simeq \eta_0 \simeq (\hbar\omega_0)^2/(6\mu_0^2 + %\pi^2T_0^2)$.
%The difference between  $\tau_0^{inter}$ and  $\tau_0^{intra}$ is primarily associated
%with a linear energy dependence of the density of states in GLs. 
In equilibrium,  Eqs.~(5) and (6) yield $T = T_0$ and ${\cal N}_{0} = {\cal N}_{0}^{eq}$.

For the rate of optical phonons decay due to the unharmonic contributions to the
interatomic potential, resulting in the phonon-phonon
scattering, one can use 
 the following simplified equation:

\begin{equation}\label{eq9}
R_0^{decay} = \frac{\Sigma_0({\cal N}_0 - {\cal N}_{0}^{eq}) }{\tau_0^{decay}},
\end{equation}
where $\tau_0^{decay}$ is the decay time of optical phonons and ${\cal N}_0 ^{eq}\simeq \exp(-\hbar\omega_0/T_0)$ is the number of optical phonons in equilibrium.
Considering high heat conductivity of GLs~\cite{32},
the lattice  temperature, i.e. the temperature
 of acoustic phonons,  is assumed to be equal to 
the temperature of the contacts $T_0$.

\section{Photocurrent and responsivity}

Using Eqs.~(4)-(9), we obtain

\begin{equation}\label{eq10}
\frac{T - T_0}{T_0}
 = \beta\,I_{\Omega}\biggl(\frac{T_0\Omega}{\hbar\omega_0^2}\biggr)
\biggl(\frac{g_{\Omega}^{intra}\eta_0}{G_0^{eq}}\biggr)\biggl(1 + \displaystyle\frac{\tau_0^{decay}}{\tau_0^{intra}}\biggr).
\end{equation}
Here we also have introduced the rate of the generation of the electron-hole pairs 
due to the absorption of equilibrium optical phonons
$G_0^{eq}  = (\eta_0\Sigma_0/\tau_0^{intra})\exp(-\hbar\omega_0/T_0)$ 
and parameter $\eta_0 = \tau_0^{intra}/\tau_0^{inter}$, where
$\tau_0^{inter}$ is the time of the interband transitions. 
The difference between $\tau_0^{intra}$ and $\tau_0^{intra}$ is due to the features
of the density of states in GLs.
At $\mu_0 \gg T_0$, one obtains~\cite{33}
$\eta_0 \simeq (\hbar\omega_0/\mu_0)^2 /6$.
The quantity $G_0^{eq}$ weakly decreases with increasing the majority carrier concentration 
(if $\mu_0 \ll \hbar\omega_0$) and strongly (exponentially) drops with decreasing temperature.
At room temperature  $G_0^{eq} \simeq (1 - 10)\times10^{20} $~cm$^{-2}$s$^{-1}$ (compare with ~\cite{24}).

One can see from Eq.~(10)  that the intraband absorption of THz radiation
leads to an obvious increase of the effective temperature $T$.

\begin{figure}[t]
\vspace*{-0.4cm}
\begin{center}
\includegraphics[width=7.0cm]{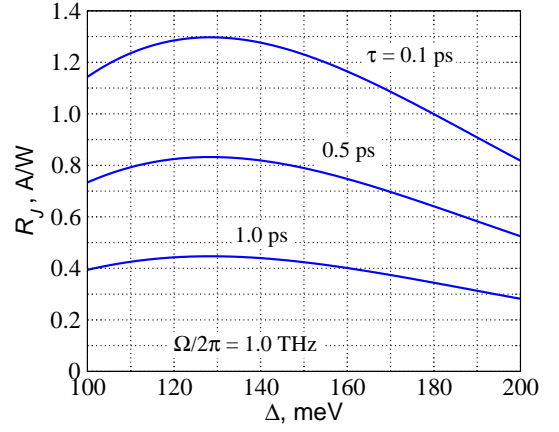}
\caption{Responsivity of the bolometers as functions
of the GNR energy gap for different $\tau$. 
%$\mu_0 = 50$meV, $\Omega/2\pi = 1.0$~THz 
}
\end{center}
\end{figure}

\begin{figure}[t]
\vspace*{-0.4cm}
\begin{center}
\includegraphics[width=7.0cm]{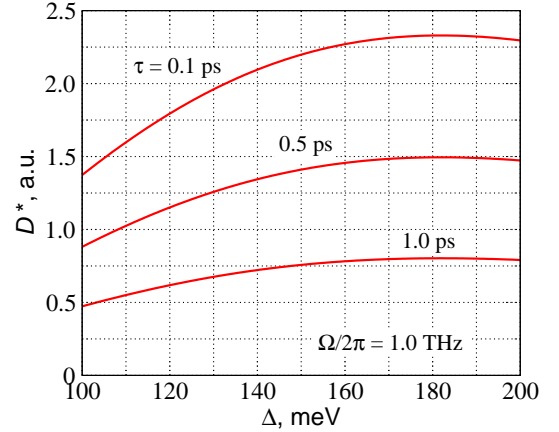}
\caption{Normalized detectivity  of the bolometers as functions
of the GNR energy gap for different $\tau$. 
%$\mu_0 = 50$meV, $\Delta = 150$~meV, $\Omega/2\pi = 0.5, 0.75, 1.0$~THz
}
\end{center}
\end{figure}

Substituting $T - T_0$ from Eq.~(10) 
into Eq.~(4), we obtain

\begin{equation}\label{eq11}
\frac{J - J_0}{N} \simeq \beta\,I_{\Omega}\biggl(\frac{4eT_0}{\pi\hbar}\biggr)\biggl(\frac{\Omega}{\omega_0}\biggr)
\biggl(\frac{T_0}{\hbar\omega_0}\biggr)\frac{\eta_0g_{\Omega}^{intra}{\cal H}}{G_0^{eq}}.
\end{equation}
Here
$$
{\cal H} = \frac{\displaystyle\exp\biggl(\frac{\mu_0 -\Delta/2}{T_0}\biggr)}{\biggl[\displaystyle\exp\biggl(\frac{\mu_0 -\Delta/2}{T_0}\biggr) + 1\biggr]}
$$
\begin{equation}\label{eq12}
\times\biggl(\frac{\Delta/2 - \mu_ 0}{T_0} + 1\biggr)\biggl(1 + \frac{\tau_0^{decay}}{\tau_0^{intra}}\biggr). 
\end{equation}
Using Eqs.~(11) and (12), for the bolometer current responsivity 
$R_J = (J - J_0)/\hbar\Omega\,I_{\Omega} S$ ($S$ is the  area of GLs), we obtain

 \begin{equation}\label{eq13}
 \frac{R_J}{N} \simeq \biggl(\frac{8\alpha\,e}{3\pi\hbar\,G_0^{eq}S}\biggr)\biggl(\frac{T_0}{\mu_0}\biggr)\biggl(\frac{T_0\tau/\hbar}{1 + \Omega^2\tau^2}\biggr) {\cal H}.
\end{equation}
For instance, considering a quasi-optic THz bolometer with a single GNR ($N = 1$) integrated with a spiral antenna, 
we can assume that $\tau = 10^{-13}$~s, $\tau_0^{decay} = \tau_0^{intra}$,     $S = 5~\mu$m$^2$, 
(about that   in ~\cite{22,23}, $T_0 = 300$~K, 
and $\Omega/2\pi = 1$~THz. Setting $\Delta = 150$~meV, $\mu_0 = 50$~meV, 
%$\mu_0 = 100$~meV  ( so that $\Sigma_0 \simeq 10^{12}$~cm$^{-2}$ 
and    $G_0^{eq} = 2.5\times 10^{20}$~cm$^{-2}$s$^{-1}$, we find $R_J \simeq 1.25$
~A/W.  If the applied bias voltage $V = 200$~mV, setting $J_0 = 2.64~\mu$A, for the voltage responsivity
$R_V = R_JV/J_0$ we obtain $R_V = 1\times 10^5 $~V/W.
The later values of the current and voltage responsivities significantly exceed those for uncooled hot-electron bolometers based on the heterostructures made of
the standard semiconductor (for example,  CdHgTe hot-electron bolometers~\cite{22}). 

Using Eqs.~(3) and (13), one can calculate the bolometer dark current limited detectivity $D^* \propto R_J/\sqrt{J_0}$. 
Since $R_J \propto N$ and $J_{0} \propto N$, $D^* \propto \sqrt{N}$ 
(for fixed $S$).
At fixed value of $\mu_0$, the detectivity achieves its maximum at 
$\Delta/2 - \mu_0 \simeq T_0$.

Equation~(12) shows that the heating of the optical phonon system due to the energy which this system receives from heated electrons and holes promotes an increase in the responsivity. 
The relative contribution of the optical phonon heating is determined by the factor $\tau_0^{decay}/\tau_0^{intra}$.
This implies that the bolometric effect in question is not purely a hot-electron or hot-hole effect. The bolometer spectral characteristic is determined by
the frequency dependence of the ac Drude conductivity, which, as seen from Eq.~(13)  at $\Omega\tau > 1$, results in $R_J \propto \Omega^{-2}$.
If $\tau = 10^{-13}$~s, the responsivity roll-off occurs at $\Omega/2\pi > 1.6$~THz.

Figures~2 and 3 show the dependences of the responsivity and detectivity, respectively,
on the energy gap in GNR, $\Delta$,  calculated for the THz bolometers with $N = 1$
and different momentum relaxation times $\tau$ for $\Omega/2\pi = 1$~THz.
It is assumed that $\mu_0 = 50$~meV and $G_0^{eq} = 3\times 10^{20}$~cm$^{-2}$s$^{-1}$.

According to Eq.~ (13), the responsivity  increases with increasing number, $N$,  of GNRs,
if the GL area $S$ is fixed. However, an increase 
in $N$ may require the related  increase in
the width of GLs and, consequently, in their area.

Similar bolometer can be based on n-GNR-n heterostructures.
The results obtained above can also be applied to this device with 
small modifications: the dark current and responsivity given by Eqs.~(3)
and (13) should be multiplied by a factor $\simeq 1/2$, because 
the terminal dark current and photocurrent are due to the electrons injected from only one
GL.

\section{Conclusion}
In conclusion,  novel   THz uncooled  bolometers based on
 nG-GNR-pG  heterostructures have been proposed.
 Using the developed model, we calculated the bolometer dark current and 
 responsivity and demonstrated that nGL-GNR-pGL can surpass the hot-electron bolometers
 based on traditional semiconductor heterostructures.

\section{Acknowledgment}
This work was supported by the Japan Society for Promotion of Science  and
TERANO-NSF grant, USA. The work at RPI  was supported by NSF and ARL Alliance Cooperative Research Agreement program.

%\newpage

\end{document}